\documentclass{aa}  

\usepackage{graphicx}
\usepackage[varg]{txfonts}
\usepackage{url}
\usepackage{color}
\usepackage{hyperref}

\begin{document} 
\newcommand\source{Swift\,J1816.7--1613}

   \title{Properties of the transient X-ray pulsar Swift\,J1816.7--1613 and its optical companion}
   \titlerunning{X-ray pulsar Swift\,J1816.7--1613}

   \author{Armin~Nabizadeh\inst{1}
          \and Sergey~S.~Tsygankov\inst{1,2} 
          \and Dmitrij~I.~Karasev\inst{2}
          \and Juhani~M{\"o}nkk{\"o}nen\inst{1}
          \and Alexander~A.~Lutovinov\inst{2}
          \and Dmitrij~I. Nagirner\inst{4}
          \and Juri~Poutanen\inst{1,2,3}
          }
   \authorrunning{A.~Nabizadeh et al. }
   
\institute{Department of Physics and Astronomy, FI-20014 University of Turku, Finland;
\email{armin.nabizadeh@utu.fi}\and
Space Research Institute, Russian Academy of Sciences, Profsoyuznaya str. 84/32, Moscow 117997, Russia 
\and Nordita, KTH Royal Institute of Technology and Stockholm University, Roslagstullsbacken 23, 10691 Stockholm, Sweden
\and Sobolev Astronomical Institute, Saint Petersburg State University, Staryj Peterhof, Saint Petersburg 198504, Russia}


 
  \abstract{We present results of investigation of the poorly studied X-ray pulsar \source\ during its transition from the type I outburst to the quiescent state. Our studies are based on the data obtained from X-ray observatories \textit{Swift}, \textit{NuSTAR} and \textit{Chandra} alongside with the latest IR data from UKIDSS/GPS and \textit{Spitzer}/GLIMPSE surveys. The aim of the work is to determine parameters of the system: the strength of the neutron star magnetic field and the distance to the source, which are required for the interpretation of the source behaviour in the framework of physically motivated models. No cyclotron absorption line was detected in the broad-band energy spectrum. However, the timing analysis hints at the typical for the X-ray pulsars magnetic field from  a few $\times 10^{11}$ to a few $\times 10^{12}$ G. We also estimated type of the IR-companion as a B0-2e star located at distance of 7--13~kpc. }


   \maketitle
%

\section{Introduction}

The transient X-ray pulsar (XRP) \source\ was first detected by the \textit{Swift}/BAT monitor on March 24, 2008 \citep{Krimm2008}. The pulsations with the period of $142.9\pm0.2$~s were discovered in the archival \textit{Chandra}/ACIS data obtained serendipitously on February 11, 2007 \citep{Halpern2008}. An analysis of two other observations of the source performed by {RXTE}/PCA on March 29 and April 7, 2008 revealed a weighted average pulse period of $143.2\pm0.1$~s \citep{krimm2013swift}, with a hint for the spin-up between these two observations with $\dot{P}=-5.93\times10^{-7}$~s~s$^{-1}$.

Based on the observed periodicity of the outbursts detected by {\it Swift}/BAT, \citet{Corbet2014} proposed an orbital period of 151.4$\pm$1.0~d for the system. However, \citet{la2014orbital} reported a different orbital period of 118.5$\pm$0.8~d using a 113-month BAT survey. Recently, \citet{corbet2017} searched for the periodicity of the system by performing power spectral analysis covering two different time intervals. First, they used the full BAT light curve of the source and found an orbital period of 151.1$\pm$0.5~d which is consistent with the one reported by \citet{Corbet2014}, but they have not detected the orbital period of 118.5~d. Secondly, \citet{corbet2017} considered the same time restrictions as \citet{la2014orbital} used in their analysis and found a period of 118.9$\pm$0.6~d. \citet{corbet2017} stated that the difference between the orbital periods is mainly because of the usage of the different time intervals and they both may be not actual orbital period of the source. 

The energy spectrum of \source\ obtained with the \textit{Chandra} observatory on February 11, 2007 was described by a simple power-law model with a photon index of $\Gamma$ = 1.2, modified by the photoelectric absorption with $N_{\rm H} = 1.2 \times 10^{23}$~cm$^{-2}$, resulting in a flux of $4 \times 10^{-12}$ erg cm$^{-2}$ s$^{-1}$ in the 2--10 keV energy band \citep{Halpern2008}. The broad-band combined {\it Swift}/XRT and BAT spectrum obtained from the outburst episodes in the energy range of 0.2--150 keV was analysed by \citet{la2014orbital}. It was modeled by a power-law model with a photon index of $\Gamma$ $\sim$ 0.1 with a high energy cutoff at $\sim$~10 keV, resulting in the average flux of about $1.3 \times 10^{-10}$~erg~cm$^{-2}$~s$^{-1}$ in the 0.2--150 keV energy range. The column density was estimated to be $N_{\rm H} = 1.0 \times 10^{23}$~cm$^{-2}$ which is significantly higher than the average Galactic value in the source direction of $1.35\times 10^{22}$~cm$^{-2}$ obtained from {\sc nhtot} \citep{Willingale2013}.\footnote{\url{http://www.swift.ac.uk/analysis/nhtot/}} The authors also derived the off-outburst flux from the source of $\approx 2.9\times 10^{-12}$ erg cm$^{-2}$ s$^{-1}$ in the same energy band. \source\ was also detected by \textit{BeppoSAX} in September 29, 1998 with a flux of $3.6 \times 10^{-11}$ erg cm$^{-2}$ s$^{-1}$ in 15--30 keV \citep{Orlandini2008} and by \textit{XMM-Newton} under the name 2XMM J181642.7--161320 on March 8, 2003 at the lowest flux of $7 \times 10^{-13}$ erg cm$^{-2}$ s$^{-1}$ in the 0.2--12 keV band \citep{Halpern2008}.

The nature of the companion star in this system has not been determined yet. The archival \textit{Swift}/UVOT observations of the source in the given \textit{Chandra} location did not show any optical counterpart. However, taking into account the pulse and orbital periods and their location in the Corbet diagram for high-mass X-ray binaries \citep{corbet1986three}, both \citet{Corbet2014} and \citet{la2014orbital} argued that the type of the system is consistent with a Be/neutron star binary. 

The magnetic field of the neutron star (NS) in \source\ has never been measured, and the properties of the system in the hard X-ray band remained unknown. In the current work we use our single {\it NuSTAR} observation in the hard X-rays as well as multiple archival observations in the standard X-ray energy band and the infrared (IR) data in order to determine properties of the system and its optical companion.  

\begin{table}
	\centering
	\caption{X-ray observations of \source.}
	\label{tab:1}
	\begin{tabular}{lccc} 
		\hline
		Satellite & Date (MJD) & Instrument & Exposure (ks)\\
		\hline
		\textit{Swift} & 54557--54578 & XRT & 10.2\\
        			   & 57803--57866 & XRT & 23.0\\
		\textit{Chandra} & 54142 & ACIS & 16.0\\
		\textit{NuSTAR} & 57807 & FPMA/B & 50.4\\
		\hline
	\end{tabular}
\end{table}  

\section{Observations and data reduction}

This study is based on the data provided by three X-ray space satellites \textit{Swift}, \textit{Chandra} and \textit{NuSTAR} and the latest public IR data from the {UKIDSS/GPS} and \textit{Spitzer}/GLIMPSE surveys. The details of the X-ray observations are given in  Table~\ref{tab:1}. For all timing and spectral analysis we use {\sc heasoft} 6.22.1\footnote{\url{http://heasarc.nasa.gov/lheasoft/}} and {\sc xspec} version 12.9.1p.\footnote{\url{https://heasarc.gsfc.nasa.gov/xanadu/xspec/manual/XspecManual.html}}

\subsection{\textit{Swift}/XRT observations}

\source\ has been observed by the XRT telescope
\citep{burrows2005swift} on-board the {\it Neil Gehrels Swift Observatory} \citep{gehrels2004swift} multiple times in 2008 and 2017 with total exposures of 10.2 and 23~ks, respectively (see  Table~\ref{tab:1}). All the {\it Swift} observations used in this study were obtained by the XRT telescope in the photon counting (PC) mode. To produce the cleaned event files (level 2) from level 1 products, the {\sc xrtpipeline} tool from the {\sc xrtdas} package v3.4.0\footnote{\url{http://www.swift.ac.uk/analysis/xrt/}} was used. Also, we used {\sc xselect} V2.4d with the standard criteria to reduce and analyze \textit{Swift}/XRT data of \source. We extracted the X-ray spectrum and the light curve of the source for every single observation in the energy range of 0.5--10~keV using circular region with a radius of 25\arcsec. The background spectra and light curves, likewise, were extracted from a source-free region with radius of 50\arcsec. The standard grade filtering (0--12 for PC) was used for the analysis. In order to calculate the ancillary response files for each observation, we used the task {\sc xrtmkarf}. 
  
 

\begin{figure}
\begin{center}    
\includegraphics[width=1.03\columnwidth]{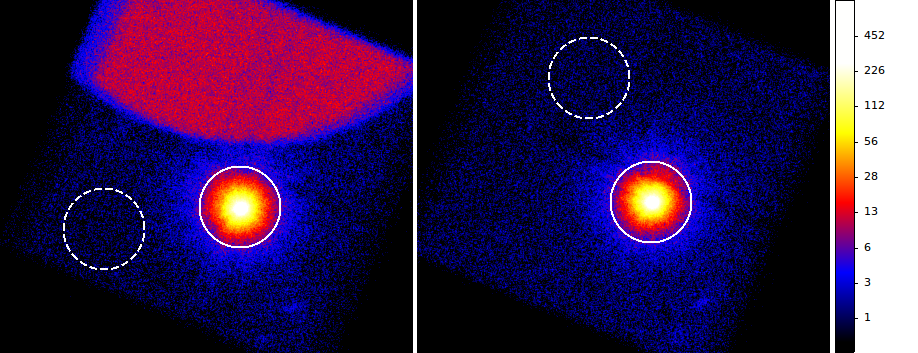}
\end{center}    
    \caption{{\it NuSTAR} image of \source\ extracted from FPMA (left) and FPMB (right) in the energy range of 3--79 keV. The solid white circles are the source extraction regions of radius 100\arcsec and the dashed white circles are the background regions of  the same radius which were selected from the source-free regions. The colour bar on the right hand side shows the number of counts per pixel.} 
    \label{fig:ds9}
\end{figure}
  
\subsection{\textit{Chandra} observations}
 
\source\ was also observed by the {\it Chandra} \citep{weisskopf1996} advanced CCD Imaging Spectrometer \citep[ACIS;][]{garmire2003} for about 16 ks on February 11, 2007 (MJD 54142; ObsID 6689). The source was placed at the ACIS-S2 with no gratings in use yielding a moderate energy resolution. To reproduce and reprocess the archival level 1 data, the standard pipeline processing was performed using the software packages {\sc ciao} v4.9 \citep{Fruscione2006} with a suitable {\sc caldb} v4.7.7. The source and background light curves and spectra were extracted from circular regions of the same radius of 20\arcsec.


\subsection{\textit{NuSTAR} observation}

The Nuclear Spectroscopic Telescope Array (\textit{NuSTAR}), launched on June 13, 2012 is the first focusing hard X-ray telescope which is operating in a wide energy range of 3--79~keV \citep{Harrison2013}. It consists of two co-aligned grazing incidence X-ray telescope systems which provides imaging resolution of  18\arcsec (full width at half-maximum, FWHM). The instruments have the spectral energy resolution of 400 eV (FWHM) at 10 keV.

In order to reduce the raw observational data, we performed the standard data reduction procedure described in the \textit{NuSTAR} user guide\footnote{\url{https://nustar.ssdc.asi.it/news.php\#}} using the \textit{NuSTAR} Data Analysis Software {\sc nustardas} v1.8.0 with a {\sc caldb} version 20180419. The source spectra and the light curves were extracted from a circular region with radius of 100\arcsec for both FMPA and FMPB using {\sc nuproducts} task (Fig.~\ref{fig:ds9}). The background photons were extracted from the source-free circular regions of the same radius located close to the source in the nearby chips because the background area in the source chip is contaminated by a faint X-ray source (bottom right corner of the source chip). Also, as shown in Fig.~\ref{fig:ds9}, a large fraction of the FPMA field of view is contaminated by stray-light. The barycentric correction was applied to the resulting light curves using standard {\sc barycor} task and the position of the source from the \textit{XMM-Newton} catalog (source XMM\,J181642.7--161320).

\begin{figure}
\begin{center}
\includegraphics[width=1.07\columnwidth]{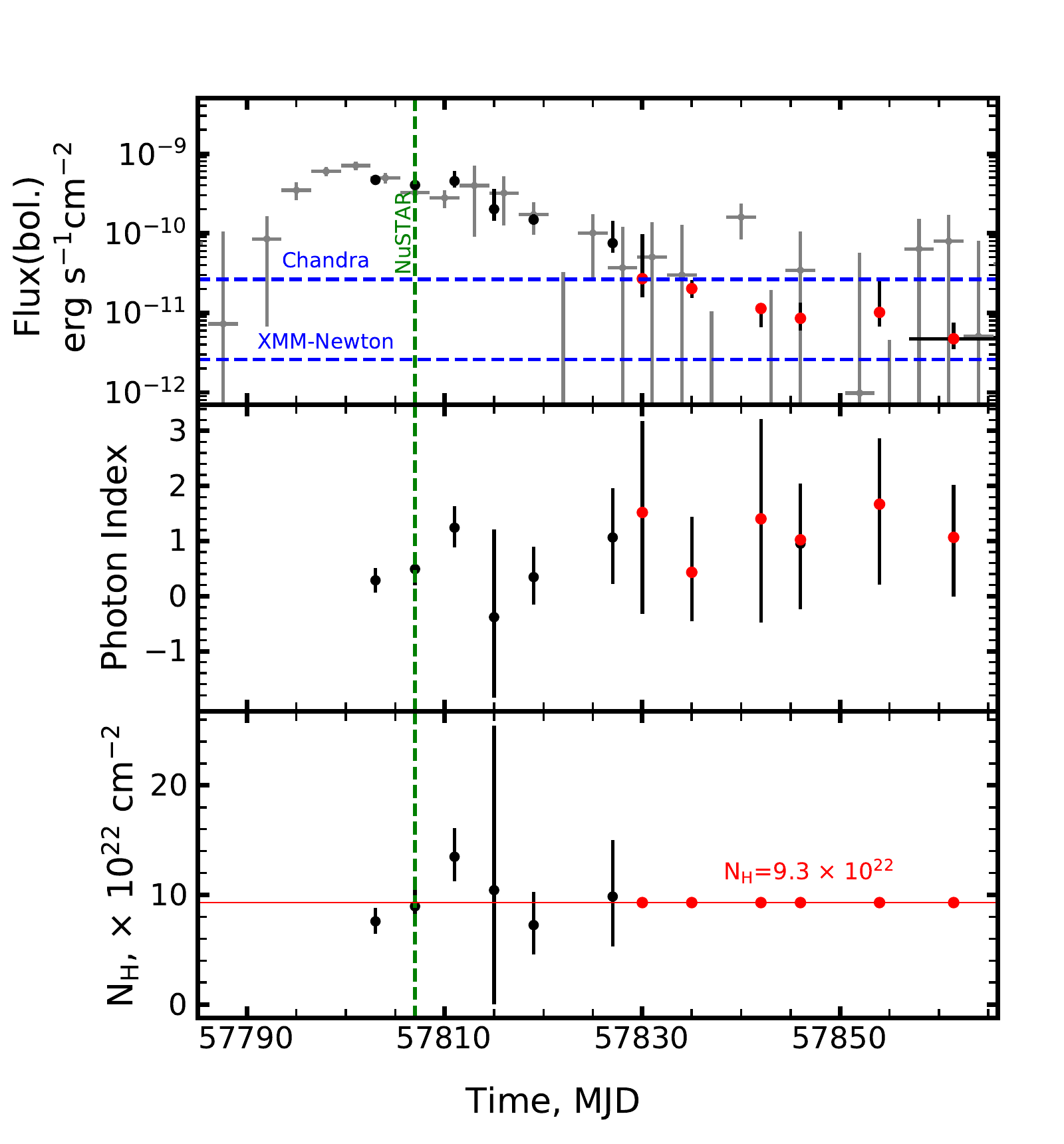}
\end{center}
\caption{\textit{Top}: Bolometric absorption-corrected X-ray light curve of \source\ obtained by \textit{Swift}/XRT in 2017. Black points correspond to individual XRT observations and the red points show the observations in which we fixed the $N_{\rm H}$ at the average value of $9.3\times 10^{22}$ cm$^{-2}$.  Gray points represent the 3-day-averaged \textit{Swift}/BAT flux in the 15--50 keV band (in arbitrary units). \textit{Middle}: Variations of the power-law photon index. \textit{Bottom}: Variations of the hydrogen column density $N_{\rm H}$. The vertical green line indicates the date of our \textit{NuSTAR} observation.} 
    \label{fig:ltc}
\end{figure}

\subsection{Infrared Observations}
\label{sec:data}

In order to study the companion star in the IR band we use the latest data of the public release of {UKIDSS/GPS}\footnote{\url{http://surveys.roe.ac.uk/wsa/}} and \textit{Spitzer}/GLIMPSE\footnote{\url{http://vizier.u-strasbg.fr/viz-bin/VizieR?-source=\%20GLIMPSE}} surveys. Based on these surveys and the catalog position of 2XMM\,J181642.7--161320 we can identify its IR-counterpart as UGPS J181642.74--161322.2 or GLIMPSE:G014.5877+00.0913 (see Table~\ref{tab:IR_Swift}). 
\begin{table}
	\centering
	\caption{Coordinates and IR-magnitudes of the counterpart of \source\ based on {UKIDSS/GPS} and \textit{Spitzer} data.}
	\label{tab:IR_Swift}
	\begin{tabular}{lc}
			\hline
	RA & 274.1781 \\
	Dec & $-$16.2229 \\ 
	$l$ & 14.5877 \\
	$b$ &  0.0913 \\
	H &  $17.56\pm0.10$  \\
	K & $14.85\pm0.02$ \\
	$[3.6]$ & $12.98\pm0.08$ \\
	$[4.5]$ & $12.52\pm0.14$  \\
	$[5.4]$ & $11.77\pm0.20$ \\ 
%
		\hline
	\end{tabular}
\end{table}

\section{Analysis and results}

In this section we present the results of the timing and spectral analysis performed on X-ray data as well as analysis of the infrared observations.

\subsection{Long-term light curve}

To investigate the long-term variability of the source we use individual XRT observations obtained in 2017. The bolometric, absorption-corrected light curve is shown in the upper panel of Fig.~\ref{fig:ltc}. 

First of all, we fitted individual spectra using a simple power law model modified with the photoelectric absorption ({\sc phabs $\times$ po} in {\sc xspec}). Due to low count rates all the spectra were binned to have at least one count in each energy channel and fitted using the W-statistics \citep{Wachter1979}. In order to find the bolometric flux from the source we converted the unabsorbed flux in the narrow energy band (0.5 -- 10 keV) to a wider energy range (0.5 -- 100 keV) using the bolometric correction factor $K_{\rm bol}={F_{\rm 0.5-100~keV}}/{F_{\rm 0.5-10~keV}}$. To calculate $K_{\rm bol}$ we used the simultaneous {\it Swift}/XRT and \textit{NuSTAR} observations. From the measured fluxes $F_{\rm 0.5-100~keV} = 4.6 \times 10^{-10}$ and $F_{\rm 0.5-10~keV}=1.3\times 10^{-10}$ erg s$^{-1}$ cm$^{-2}$ (see Sect.~\ref{sec:spectra}), we estimated the bolometric correction factor $K_{\rm bol}\approx 3.7$. To find the X-ray flux in the 0.5--100 keV energy band, the spectral model (see Sect.\,3.4) was extrapolated to 100 keV. We finally note that in our analysis the value of $K_{\rm bol}$ was assumed to be independent of the source flux.

The variability of the photon index and hydrogen column density calculated in the 0.5--10 keV band are shown in the middle and bottom panels of Fig.~\ref{fig:ltc}, respectively. For the three data points, shown with red dots, we fixed $N_{\rm H}$ to the average value of $9.3 \times 10^{22}$ cm$^{-2}$ because of the low-count statistics. The last data point (around MJD 57860) presents the parameters obtained from the averaging of two nearby observations. The blue horizontal dashed lines show the source fluxes obtained from \textit{Chandra} (MJD 54142) and \textit{XMM-Newton} (MJD 52706) observations of $7.1 \times 10^{-12}$ and $7 \times 10^{-13}$ erg s$^{-1}$ cm$^{-2}$ in the 2--10 keV energy band, respectively. For both possible values of the orbital period of the system of 118.5 and 151.1~d, \textit{Chandra} and \textit{XMM-Newton} observations correspond to the different orbital phases not close to the periastron. These fluxes as well as the flux $2.9 \times 10^{-12}$ erg s$^{-1}$ cm$^{-2}$ obtained from XRT and BAT observations during the off-outburst state of the source \citep{la2014orbital}, suggest that flux never drops much below $\sim 10^{-12}$ erg s$^{-1}$ cm$^{-2}$. The green dashed vertical line indicates the date of our \textit{NuSTAR} observation. 

As can be seen from Fig.~\ref{fig:ltc} the peak of the outburst in both XRT and BAT data is reached around MJD~57800. It is worth mentioning that this date deviates  by few tens of days from the expectations derived using orbital period and phasing reported by \citet{la2014orbital} and \citet{corbet2017}.
Most probably this is because both values of the orbital period of the source do not represent its true value due to non-periodic behaviour of the source as discussed in \citet{corbet2017}.


\begin{figure}
\begin{center}    
\includegraphics[width=1.1\columnwidth]{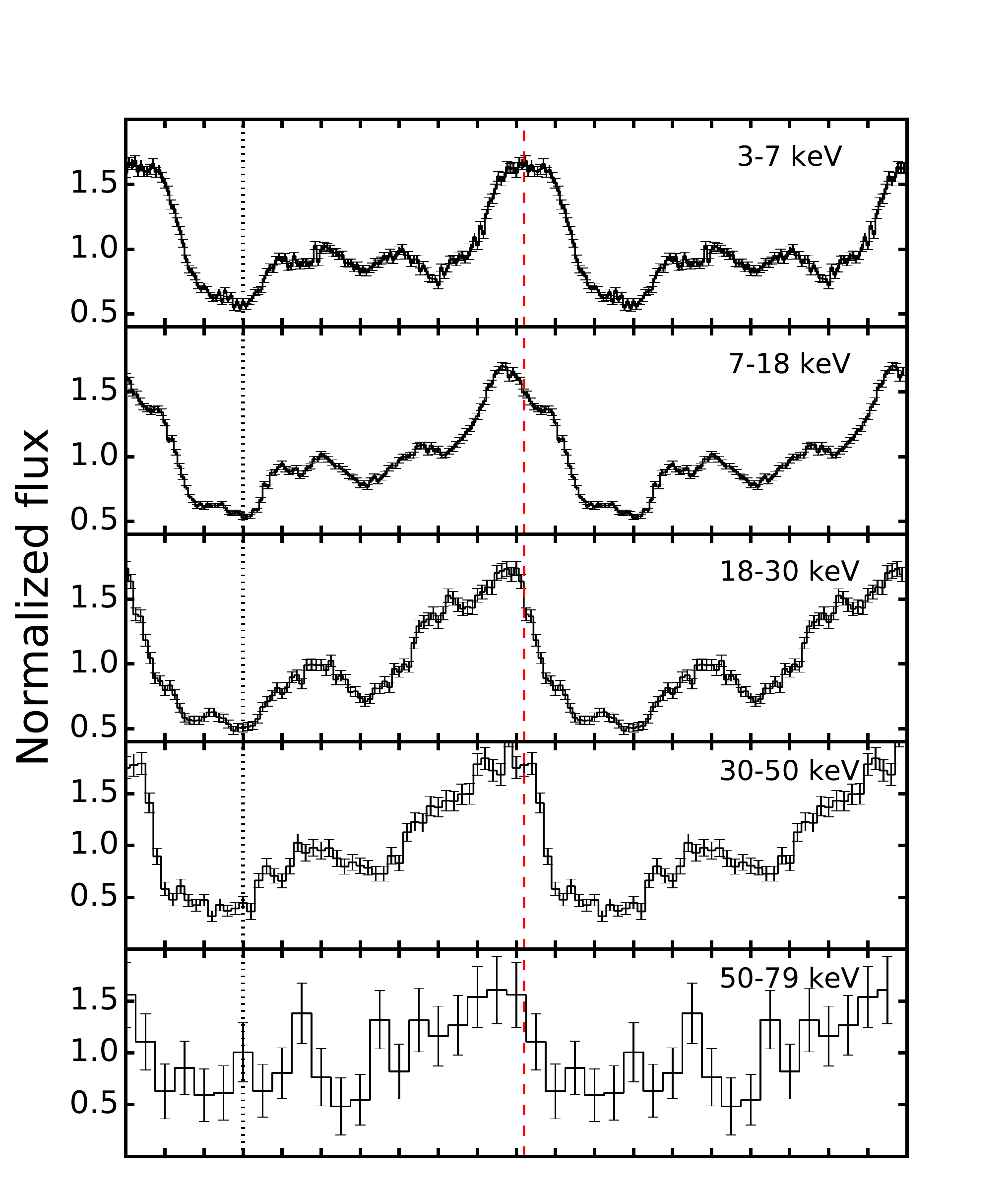}
\end{center}
\vspace{-1.2cm}
\begin{center}    
\includegraphics[width=1.1\columnwidth]{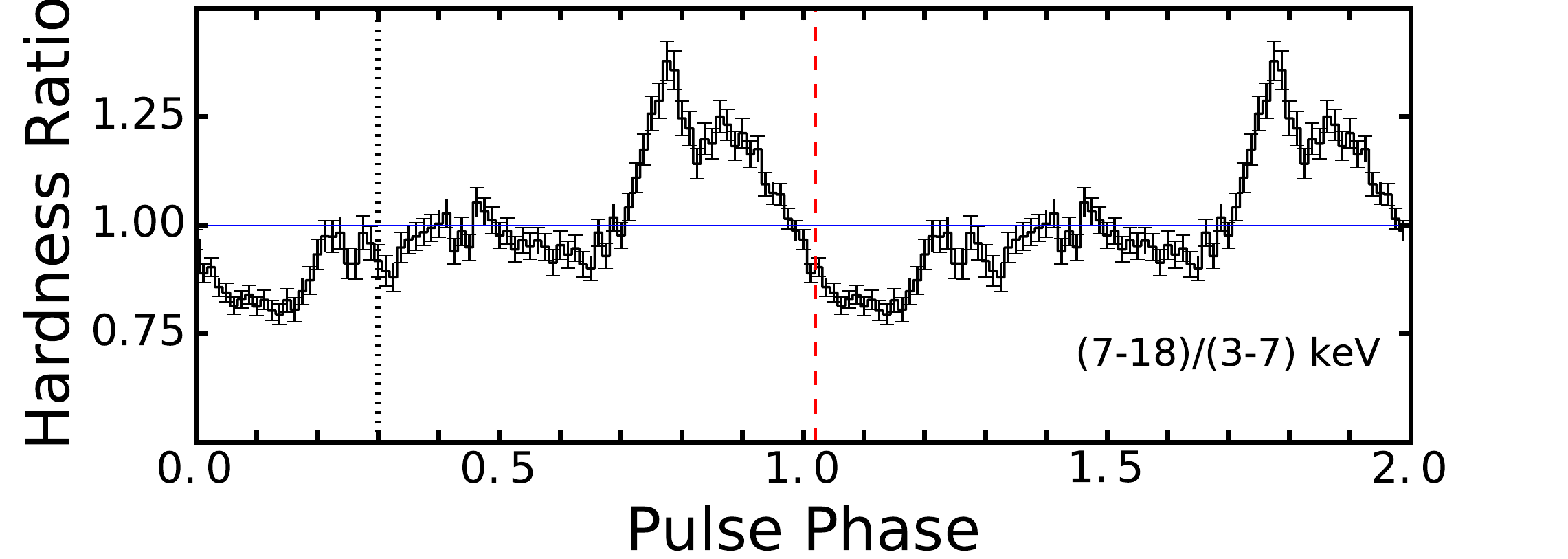}
\end{center}
\caption{\textit{Top panels}: Pulse profiles of \source\ in five different energy bands 3--7, 7--18, 18--30, 30--50 and 50--79 keV (from top to bottom) obtained with the \textit{NuSTAR} observatory. The fluxes are normalized by the mean flux in each band. The position of the main maximum and minimum are shown by vertical dashed and dotted lines, respectively. The zero phase was chosen arbitrarily. \textit{Bottom panel}: Hardness ratio of \source\ pulse profiles in the energy bands 7--18/3--7 keV. The blue horizontal line indicates the hardness ratio of unity.} 
\label{fig:pp} 
\end{figure}

\subsection{Pulse profile and pulsed fraction}

Because no accurate orbital parameters required for the correction for the binary motion are available for the source, we used only the barycentric-corrected light curves in our temporal analysis. Therefore, the yielding periods might be affected by the orbital motion of the system. Using our \textit{NuSTAR} data we obtained the spin period $P_{\rm spin}$ = 143.6863(2)~s which is consistent within errors with the value obtained by \citet{la2014orbital}. The spin period and its uncertainty were obtained from the distribution of the results of the period search in 10$^{3}$ simulated light curves following procedure described by \cite{Boldin2013}. To get the pulse period in each light curve, we use the standard {\sc efsearch} procedure from the {\sc ftools} package.

Thanks to the wide energy coverage of the \textit{NuSTAR} observatory, we can study the pulse profile dependence on energy. We first extracted the barycentric-corrected light curves in five energy bands  3--7, 7--18, 18--30, 30--50 and 50--79 keV and then combined the individual light curves from modules FPMA and FPMB to get better statistics \citep[see details in][]{Krivonos2015}. 

Using program {\sc efold}, the obtained light curves were folded with the pulse period to get the pulse profile in each energy band. Five top panels in Fig.~\ref{fig:pp} show the evolution of the pulse profile with increasing energy (from top to bottom). The pulse profiles have quite complicated structure with multiple peaks. The dominating features -- the main maximum around phases 0.9--1.1 and the main minimum around phases 0.2--0.3 (zero phase was chosen arbitrarily) -- are shown by the dashed and dotted lines, respectively. The pulse profile demonstrates a clear dependence on energy. For instance, the right part of the main peak disappears at higher energies. To illustrate the energy dependence, we plotted the hardness ratio of the fluxes in the 7--18 and 3--7 keV bands in the bottom panel of Fig.~\ref{fig:pp}. The described behaviour is clearly seen from the obvious decrease of the hardness ratio around the phase 0.1 (1.1). From another hand there is a clear hardening of the emission at phases 0.7--0.9 (the left wing of the main peak).

\begin{figure}
\begin{center}    
\includegraphics[width=1.1\columnwidth]{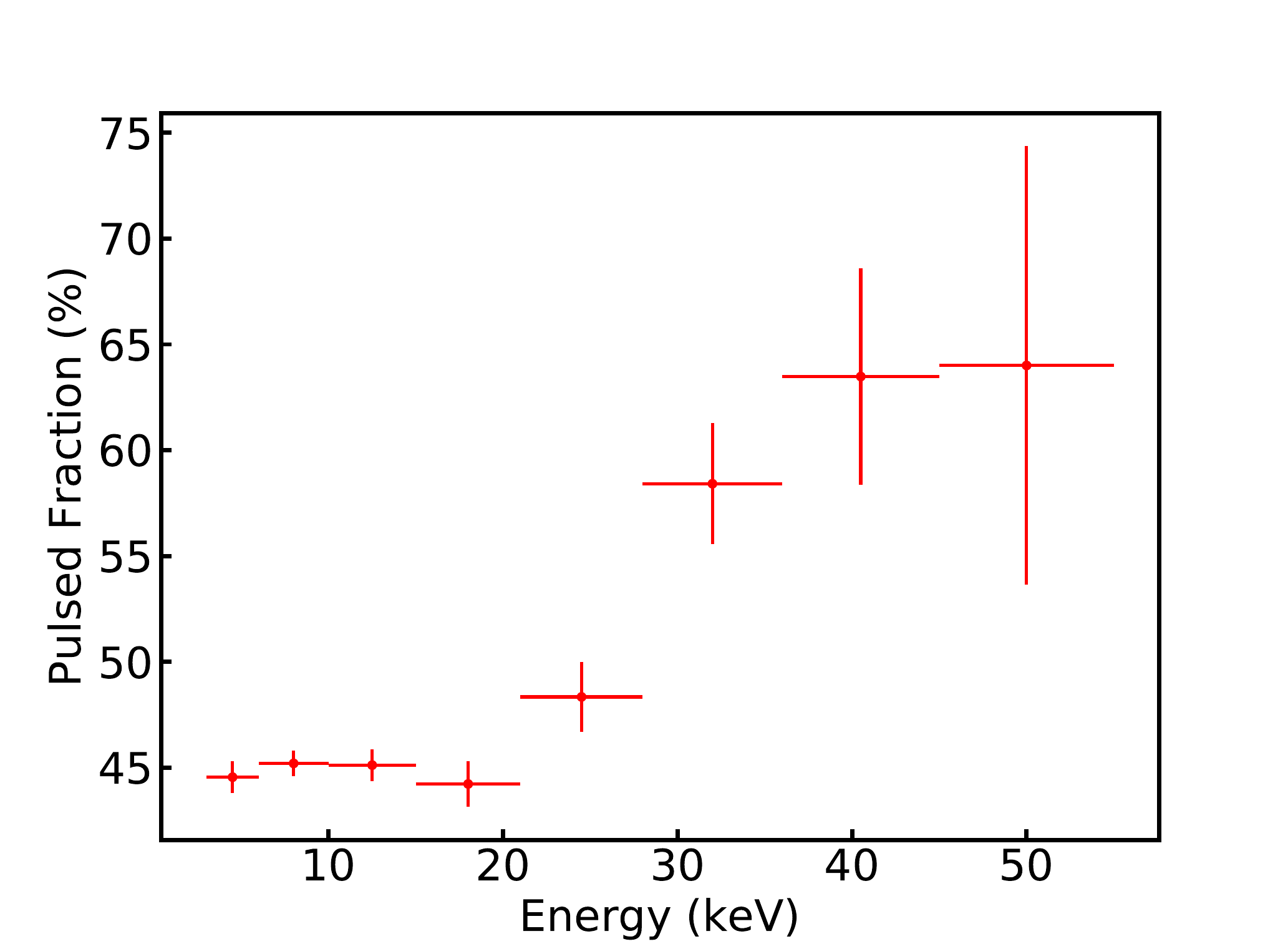}
\end{center}    
\caption{The dependence of the pulsed fraction of \source\ on energy obtained from the \textit{NuSTAR} observation.} 
    \label{fig:pf}
\end{figure}

The pulsed fraction calculated as PF = $(F_{\rm max} - F_{\rm min})/(F_{\rm max} + F_{\rm min})$ is plotted as a function of energy in Fig.~\ref{fig:pf}. $F_{\rm min}$ and $F_{\rm max}$ are the minimum and maximum fluxes of the pulse profile, respectively. Below $\sim$30 keV the pulsed fraction has a nearly constant value around 45--50\%. However, at the energies above 30 keV, it increases above 60\% that is typical for most of the XRPs \citep{Lutovinov2009}.

\subsection{Power density spectrum}

\begin{figure}
\begin{center}
\includegraphics[width=1.05\columnwidth]{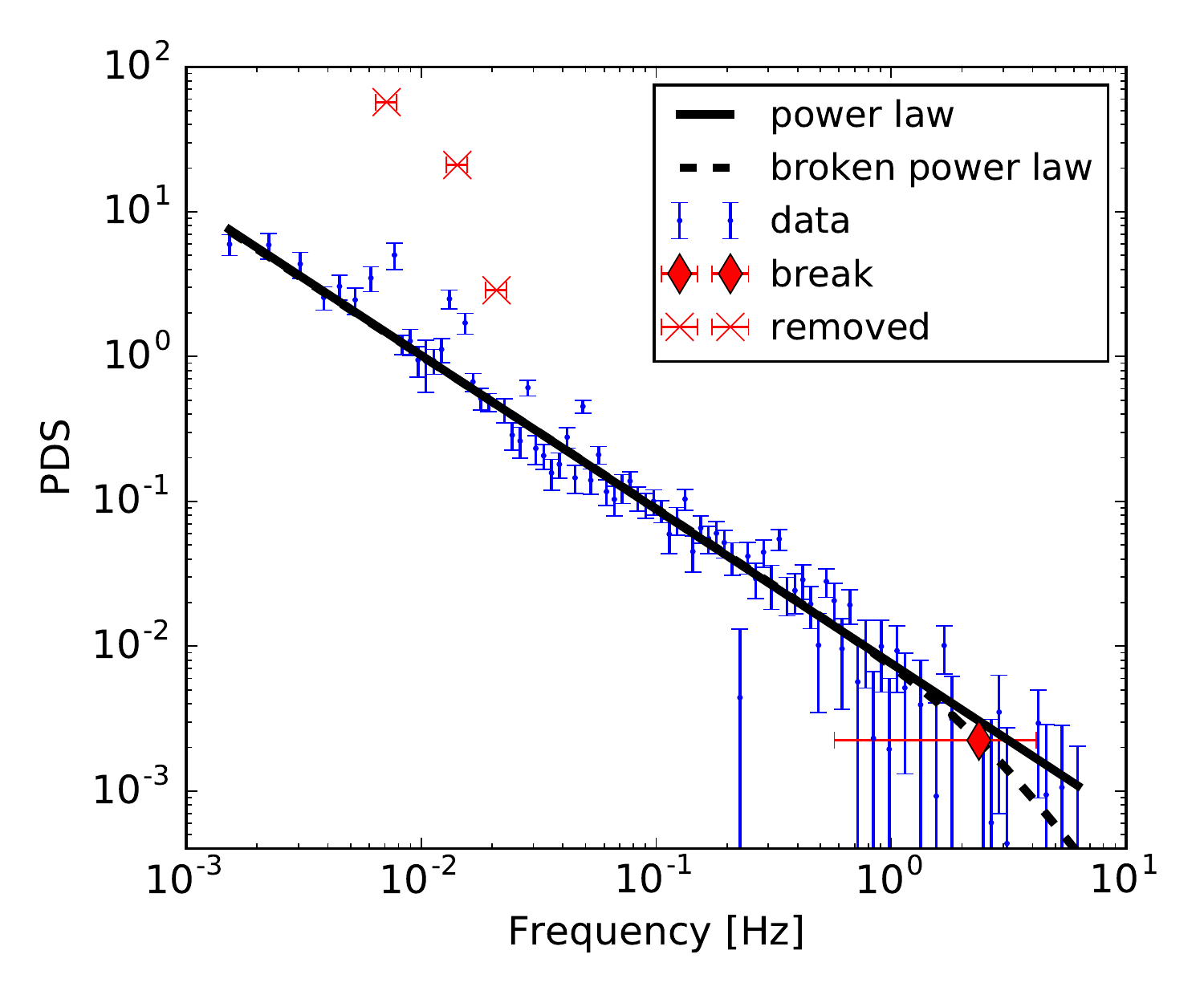}
\end{center}
\caption{Power density spectrum of \source\ obtained from combined light curves from FPMA and FPMB in the 3--79 keV energy band.  Solid line represents a single power-law fit with index $-1.1$.  
Regular pulsations at $\sim$7~mHz and their harmonics were masked out in the fit. 
The dashed line is the broken power-law fit and the red diamond is the resulting break frequency at $2.4\pm1.8$~Hz. } 
\label{fig:pds}
\end{figure}

Fig.~\ref{fig:pds} shows the power density spectrum (PDS) of the combined light curves from FPMA and FPMB in the full energy band. The PDS was calculated with the {\sc heasoft} software task {\sc powspec} using a Miyamoto normalization. The Poisson noise level was estimated from the high-frequency 20--100 Hz part of the PDS  and  subtracted. The continuum of the PDS was fit with two models, a single power law and a broken power law. The regular pulsations at $\approx 143$~s (0.007 Hz) and several harmonics are clearly visible in the PDS and were masked out for the fitting purposes.  The simple power law with an index of $-1.1$ is able to fit the data very well. The broken power law deviates from the single power law only above $\approx 1$~Hz and gives a poorly constrained break frequency of $2.4\pm1.8$~Hz but does not significantly improve the quality of the fit. Therefore we do not claim a detection of the break in the PDS. Rather, we take $0.6$~Hz as a lower limit for the break frequency in order to estimate the NS magnetic field strength (see Sect.~\ref{sec:discussion}).

\begin{table}
	\caption{Best fit parameters for the joint \textit{NuSTAR} and \textit{Swift}/XRT spectrum approximation with {\sc constant $\times$ phabs(comptt + gaussian)} model.}
	
	\label{tab:2}
	\begin{tabular}{lccr} 
		\hline
		Model & Parameters & Unit & Value\\
		\hline
		Constant & 	{\it NuSTAR}$^{a}$	&	& 0.968 $\pm$ 0.003 \\
        		 &	{\it Swift}/XRT$^{b}$ &	& 0.63 $\pm$ 0.02	\\
        \hline
        Phabs &  $N_{\rm H}$ & $10^{22}$ cm$^{-2}$ &  6.3 $\pm$ 0.5 \\
		\hline
		CompTT&  $T_0$ &	keV	&  1.17 $\pm$ 0.02 \\
				&  $kT$ & keV	&  6.82 $\pm$ 0.04 \\
                &  $\tau_{\rm p}$ &	&  5.77 $\pm$ 0.06 \\
		\hline
        Gaussian&  $E_{\rm Fe}$ &	keV	&  6.44 $\pm$ 0.03 \\
				&  $\sigma_{\rm Fe}$ & keV	&  0.21 $\pm$ 0.04 \\
                &  norm & $10^{-4}$ ph s$^{-1}$cm$^{-2}$ &  1.0 $\pm$ 0.1 \\
        \hline
        &  $F_{0.5-79}$	& $10^{-10}$ erg s$^{-1}$cm$^{-2}$ & 4.65 $\pm$ 0.03\\
		& $F_{0.5-10}$ &  $10^{-10}$ erg s$^{-1}$cm$^{-2}$ & 1.26 $\pm$ 0.02 \\
		\hline
		 C-value  & 	&	& 1477  \\
		 d.o.f. &	&	& 1572  \\
		 \hline
		 
	\end{tabular}
      \small
     $^{a}$Cross-calibration normalization factor between FPMA and FMPB instruments on-board \textit{NuSTAR}. $^{b}$Cross-calibration normalization factor between \textit{Swift}/XRT and \textit{NuSTAR}/FPMA instruments.
\end{table}

\begin{figure}
\begin{center}
\includegraphics[width=1.1\columnwidth]{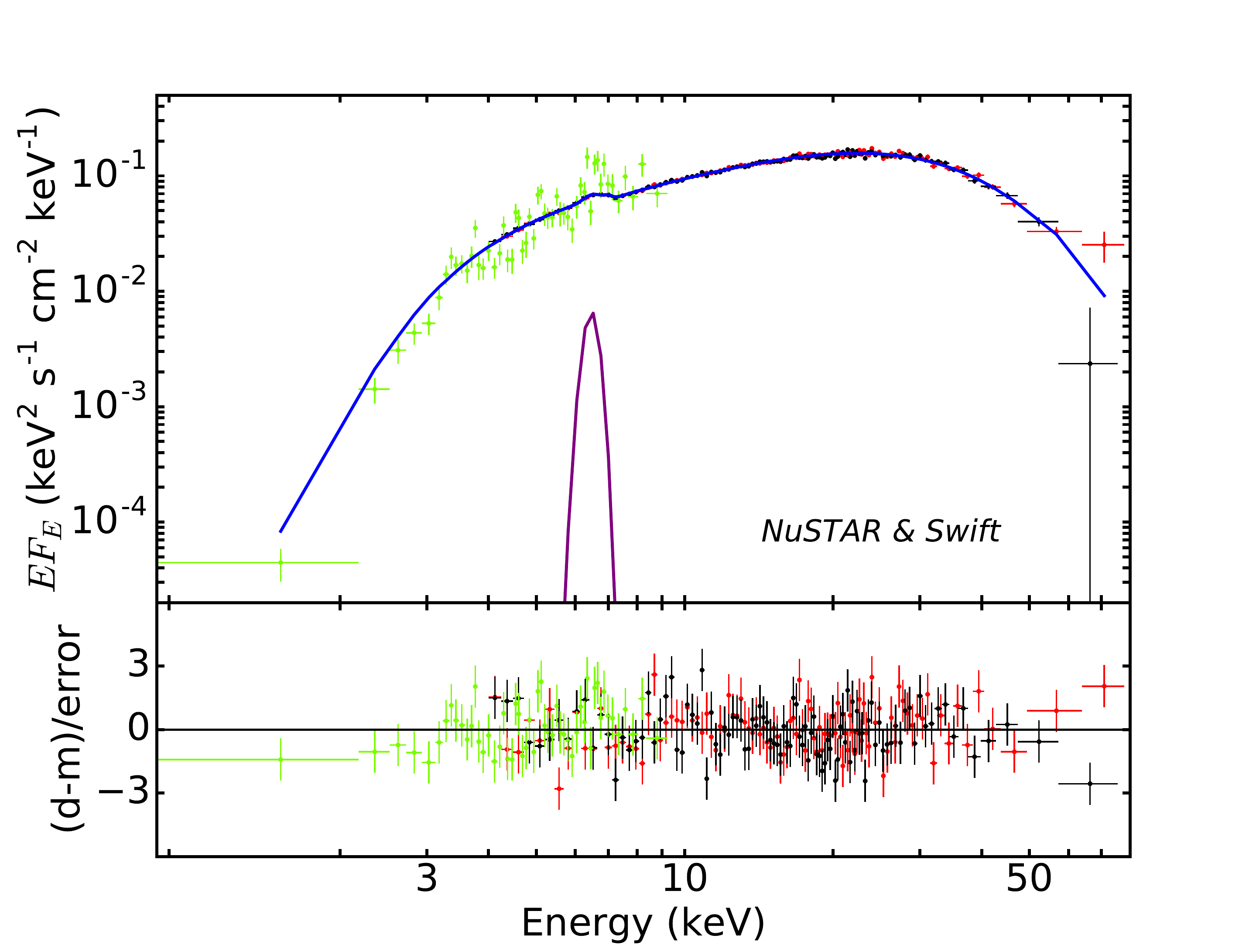}
\end{center}
\caption{\textit{Top:} the broad-band spectrum of \source\ obtained by \textit{NuSTAR}/FPMA and FPMB (red and black crosses) and \textit{Swift}/XRT (green crosses) together with the best-fit model {\sc constant $\times$ phabs(comptt + gaussian)} (solid line). \textit{Bottom:} residuals from the best-fit model in units of standard deviations.
} 
    \label{fig:spec}
\end{figure}
 
\subsection{Spectral analysis}
\label{sec:spectra}

One of the main goals of this study was searching for the cyclotron resonant scattering feature (cyclotron absorption line) in the \source\ spectrum using the \textit{NuSTAR} data. \textit{NuSTAR} presents a unique sensitivity in the energy range where most of the known cyclotron absorption feature were detected so far \citep[see][for a review]{Walter2015}. First, we extracted the spectra using the described procedure from the \textit{NuSTAR} and \textit{Swift} observations  performed simultaneously (\textit{Swift}/XRT ObsID 00031188007). It allowed us to obtain and analyse in {\sc xspec} a broad-band, 0.5--79 keV, spectrum of \source\ for the first time.

The spectrum of the source has a  shape typical for other accreting XRPs, demonstrating a cutoff at high energies \citep[see, e.g.,][]{2005AstL...31..729F}. Although, the \textit{NuSTAR} observation provided good statistics, we did not detect a cyclotron absorption line in the spectrum. To describe the spectral shape, we first fitted the  spectrum using a cut-off power-law model modified by the photoelectric absorption {\sc phabs $\times$ cutoffpl}. This is the same model as was used by \cite{la2014orbital} to fit successfully the joint XRT and BAT spectrum of the source. However, for the better quality \textit{NuSTAR} data it resulted in unacceptable C-statistic value of 2006 (for 1576 d.o.f.). A better fit quality was obtained by using the Comptonization model of soft photons in a hot plasma \citep{Titarchuk1994}, modified by the fluorescent iron line at 6.4 keV and photoelectric absorption {\sc phabs (comptt + gau)}. This composite model yielded a much lower C-value = 1477 (for 1572 d.o.f.).

The broad-band unfolded spectrum and the corresponding residuals from the best-fit model are shown in Fig.~\ref{fig:spec}. In order to take into account the uncertainty in cross-calibration between different instruments we added the normalization factor to the model. It gives a normalization of 0.97 between \textit{NuSTAR} instruments (FMPA and FMPB) and 0.63 between FMPA and \textit{Swift}/XRT. The {\sc gaussian} component representing an iron 
fluorescent emission line has peak at 6.44 keV and width of 0.21 keV. The best-fit parameters and their uncertainties are listed in Table.~\ref{tab:2}. 

We note that the best-fit value for the hydrogen column density $N_{\rm H} = (6.3\pm0.5)\times10^{22}$ cm$^{-2}$ is significantly higher than the Galactic mean value in this direction of $1.35 \times 10^{22}$ cm$^{-2}$ \citep{Willingale2013}.  Such a difference can be due to either a significant intrinsic absorption in the system or some non-uniformities in the interstellar medium in the direction to the source, which cannot be detected on the standard maps due to their angular resolution limits. The absorption value obtained in our analysis is somewhat lower than the value  $N_{\rm H} = (10.2\pm0.5)\times10^{22}$ cm$^{-2}$ calculated by \citet{la2014orbital} in the direction to the source. This discrepancy can be caused by different spectral continuum models used in the mentioned work. For the same reason $N_{\rm H}$ obtained from the broad-band spectrum is lower than that derived from the individual \textit{Swift}/XRT spectra (see the bottom panel of Fig.~\ref{fig:ltc}).

\begin{figure*}
\includegraphics[width=\columnwidth, trim={1cm 7cm 0cm 2cm},clip]{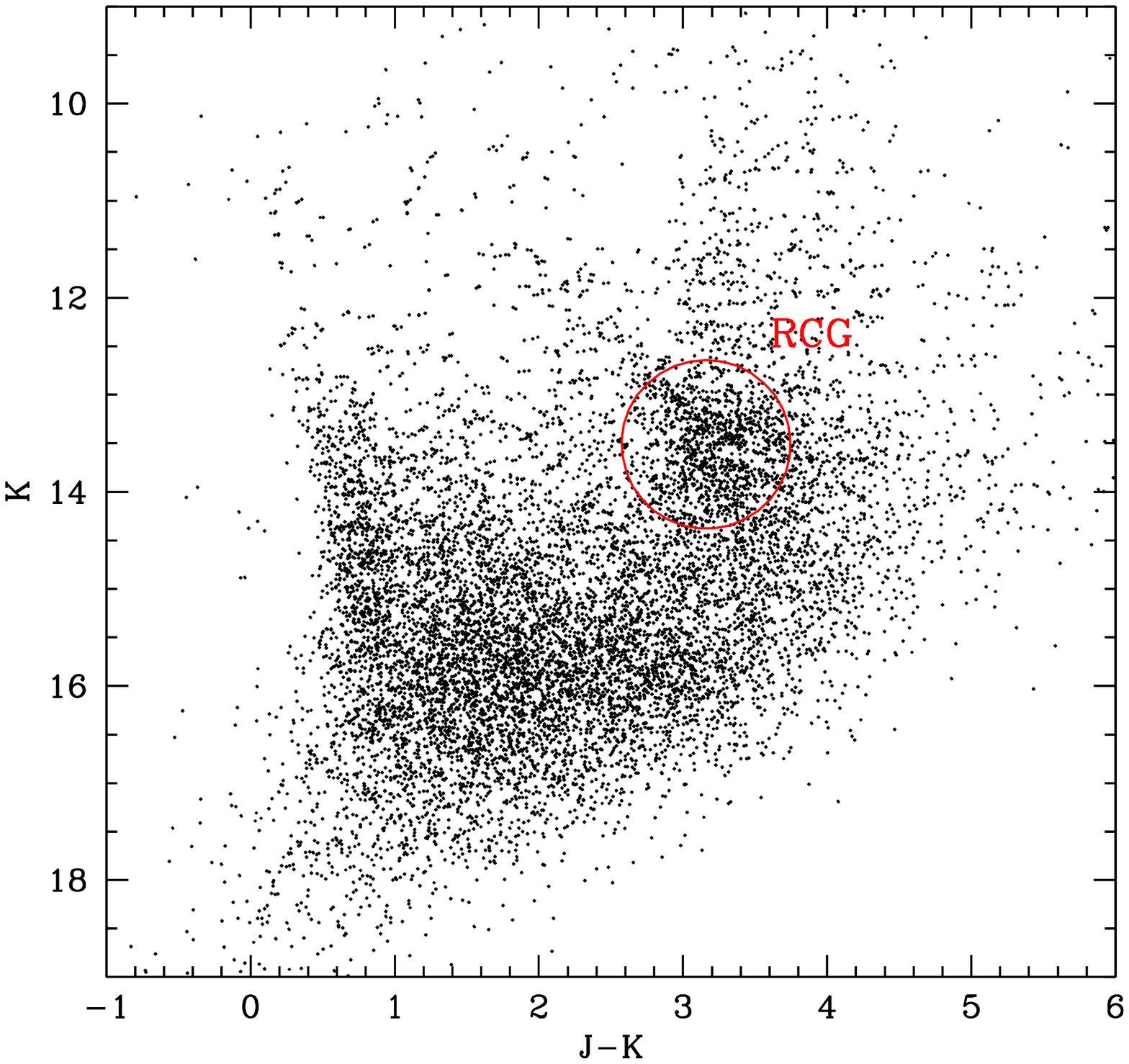} 	\includegraphics[width=\columnwidth, trim={1cm 7cm 0cm 2cm},clip]{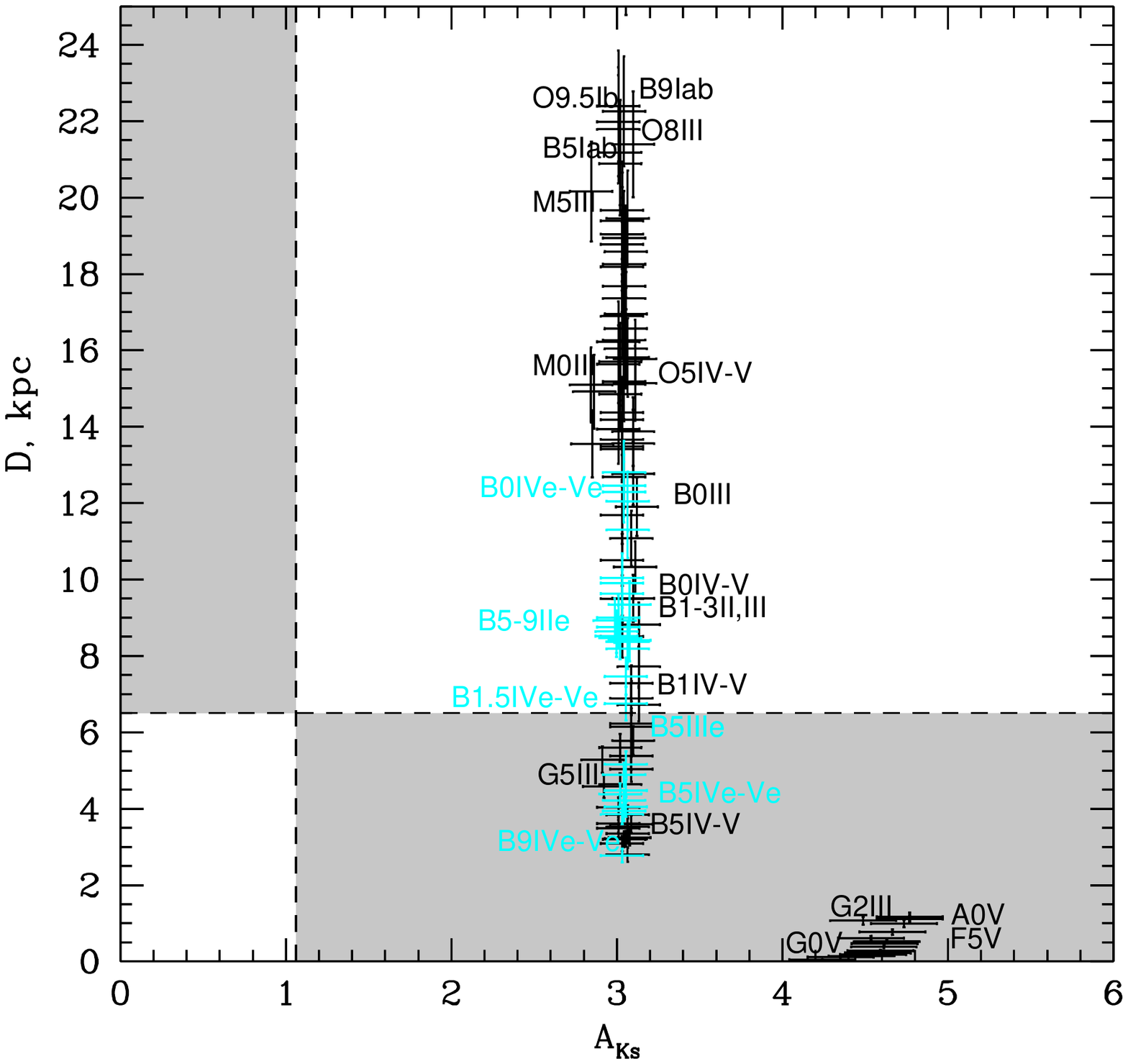}
\caption{\textit{Left panel:} Colour-magnitude diagram for all stars  from UKIDSS/GPS sky survey (see Sect.~\ref{sec:data}) in the 3\arcmin$\times$3\arcmin\ 
vicinity  around \source. Red clump giants are marked by the red circle. \textit{Right panel:} `Distance-absorption' diagram that demonstrates which types of stars can be appropriate as counterparts for \source\ (white areas). The stars (shown with 
cian crosses for Be stars and black crosses for other stars) located in the gray areas of the diagram cannot be counterparts (see text for details). The black dashed lines represent the absorption and distance to the Galactic bulge. } \label{fig:cmd}
\end{figure*}

\subsection{Identification of the IR-companion}
\label{sec:opt}

As was mentioned in Sect.~\ref{sec:spectra}, the absorption value measured from the X-ray spectrum exceeds significantly the Galactic value in the source direction.
To determine the nature of \source, we have analyzed IR properties of the counterpart according to the method proposed by \citet{Karasev2010, Karasev2015}. The idea is to get photometric observations of an object at least in two filters, as well as to know the distance and the absorption up to the Galactic bulge in the direction of the studied object. After that we can evaluate the absorption magnitude to the source and roughly estimate the possible distance and the spectral class of the companion star.
 
To obtain reliable estimates of the absorption and other parameters using this method, we need to know a correct extinction law in the direction towards the source. Here we use a non-standard law $ A_{\rm Ks}/E(J-Ks)\approx 0.43$ and $A_{\rm Ks}/E(H-Ks)\approx 1.12 $ obtained for the Galactic bulge interstellar medium by \citet{Karasev2015}, \citet{2017ApJ...849L..13A}, and \citet{2018AstL...44..220K}. It is significantly different from the standard one \citep{1989ApJ...345..245C} and more suitable for studying the central parts of the Galaxy.
 
In the next step, we are able to determine the magnitude of the absorption to the Galactic bulge using the position of the red clump giants (RCG) on the colour-magnitude diagram (CMD), reconstructed for all stars in a $\sim 3\arcmin\times 3\arcmin$ vicinity around the object (Fig.~\ref{fig:cmd}, left panel). Note that all RCG stars have approximately the same luminosity and colour (they are weakly affected by the metallicity and the age). They form a compact and well-recognized clump on the red giant branch with the observed magnitude and colour of their centroid, $Ks_{\rm RCG} =
13.51 \pm 0.05 $ and $(J-Ks)_{\rm RCG}= 3.16 \pm 0.04 $. Taking into account an absolute magnitude of RCG $M_{\rm Ks, RCG}\approx -1.61$ and their intrinsic colour $(J-Ks)_0\approx 0.68$   \citep{2000ApJ...539..732A,Karasev2010, 2012A&A...543A..13G, 2017AstL...43..545G} we can estimate the extinction and the distance to these objects (and consequently to the bulge) as $A_{\rm Ks}=1.06\pm0.03$ and
$D=6.5\pm0.2$ kpc (here we used the relations $A_{\rm J}-A_{\rm Ks} =
(J-Ks)_{\rm RGC}-(J-Ks)_0$, $A_{\rm J}\approx 3.33 \times A_{\rm Ks}$ in accordance with the above extinction law as well as the relation $5-5\log D=M_{\rm Ks, RCG} - Ks_{\rm RCG} + A_{\rm Ks}$, where $D$ is the distance in pc).\footnote{The absolute magnitude of RGC was previously converted to the UKIRT system using transformation formulas from
\url{http://www.astro.caltech.edu/~jmc/2mass/v3/transformations/}} Note that the obtained distance to the bulge at $l\approx 14\degr$ agrees roughly with the extrapolation of the results of \citet{2012ApJ...744L...8G} to these Galactic longitudes.
 
Once we determined the distance and absorption to the Galactic bulge in the source direction, we can proceed to estimate the absorption to the source as well as its class and the distance. Assuming stars of different spectral and luminosity classes as a counterpart of \source\ we can define a correction for the absorption and distance required for each of them to satisfy the observed magnitudes.\footnote{Absolute magnitudes and intrinsic colours of stars of different spectral and luminosity classes for the corresponding filters were taken from \citet{2000MNRAS.319..771W, 2006MNRAS.371..185W, 2007MNRAS.374.1549W, 2014AcA....64..261W} and \citet{2015AN....336..159W} for Be-stars.} As a result we obtain the `distance-absorption' diagram (Fig.~\ref{fig:cmd}, right panel), where classes of stars, that potentially could be a companion of \source, are located in white areas, i.e.  if the star is located in front of the bulge, the absorption to it should be lower than the absorption to the bulge, and vice versa.  Stars, which cannot be a companion are located in gray areas. Dashed lines correspond to the absorption and distance to the Galactic bulge and divide the diagram into the above mentioned areas.  Note, that we do not know precisely where the extinction law becomes a non-standard one, therefore for stars located in front of the bulge we use a standard extinction law, and for stars in the bulge or behind it the non-standard law is applied \citep[for a detailed description of the method, see][]{2015AstL...41..394K}.

From the distance-absorption diagram we can make the first essential conclusion: the absorption magnitude to \source\ is $A_{\rm Ks} \approx 3.0$. Additionally, the source counterpart belongs to the stars with classes not later than B2-3 for main sequence or Be-stars, which are located at distances more than $\approx$7 kpc (Fig.~\ref{fig:cmd}, right panel). At the same time from the source behaviour in the X-rays and the Corbet diagram, we have indirect indications that the companion of \source\ is probably a Be-star \citep{Corbet2014, la2014orbital}. From Fig.~\ref{fig:cmd} we can see that stars of B0-2e or early classes located at the distances of 7--13 kpc can be appropriate companions. We note that we are not able to relate the absorption measured in the X-rays towards the source ($N_{\rm H}$) to the magnitude of $A_{\rm Ks}$ using standard relations \citep[see e.g.][]{2009MNRAS.400.2050G}, because the extinction law towards the source is not the standard one (see above). Thus, we are unable to compare the IR and X-ray results directly and dedicated spectroscopic observations in the near IR band are required to firmly establish the nature of the companion star in \source.

\section{Discussion} 
\label{sec:discussion}
 
In this work, we presented a detailed investigation of the transient XRP \source\ both in X-ray and infrared bands. One of main goals was to estimate the fundamental physical parameters of the source such as the magnetic field of the NS. The most straightforward way to determine the magnetic field in such systems is to find the cyclotron absorption line in its spectrum. No such a feature was detected through the spectral analysis of the {\it NuSTAR} data. The lack of the cyclotron absorption line suggests that the magnetic field in \source\ is either weaker than $\sim5\times10^{11}$ G  or stronger than $\sim6\times10^{12}$~G considering the lower and upper limits of the \textit{NuSTAR} energy range where sensitivity allows us to exclude presence of such feature. 

In the literature one can find another, but indirect, methods to estimate the strength of the magnetic field in XRPs. However, all such models require a knowledge of the mass accretion rate onto the NS, and hence the distance to the system. For the following discussion we use the distance $D=10$~kpc obtained in Sect.~\ref{sec:opt}.

One of indirect methods to estimate the magnetic field is to use the so-called propeller effect  \citep{Illarionov1975,1986ApJ...308..669S}. In the propeller regime, the accretion would be prohibited by the centrifugal barrier caused by the rotating magnetosphere of the NS if the magnetospheric radius is larger than the corotation radius ($R_{\rm m} > R_{\rm c}$). It happens at the critical luminosity $L_{\rm prop}$  which is a function of the pulse period and magnetic field strength \citep[see e.g.,][]{2002ApJ...580..389C}. As a result of the propeller effect, the source luminosity drops sharply by about two orders of magnitude for a typical XRP. The method of the magnetic field estimate based on the measurement of the critical luminosity $L_{\rm prop}$ was recently calibrated using XRPs with known magnetic fields \citep{Tsygankov2016a, 2017ApJ...834..209L}. In order to detect the propeller effect, we organized a {\it Swift}/XRT monitoring of \source, however, no sharp drop of the flux was found in the tail of the studied outburst (Fig.~\ref{fig:ltc}).  

Instead of the transition to the propeller regime, the source light curve shows a slowdown of the flux decline at the level of a few $\times 10^{-12}$ erg s$^{-1}$ cm$^{-2}$ (see top panel of Fig.~\ref{fig:ltc}), that corresponds to the source luminosity of a few $\times 10^{34}$ erg s$^{-1}$. Although, the XRT monitoring was not long enough to cover full orbital cycle, the source fluxes measured by \textit{Chandra} and \textit{XMM-Newton} ($2.6 \times 10^{-11}$ and $2.6\times 10^{-12}$ erg s$^{-1}$ cm$^{-2}$ after bolometric correction, respectively) as well as the off-outburst flux of $2.9 \times 10^{-12}$~erg s$^{-1}$ cm$^{-2}$ obtained by XRT and BAT \citep{la2014orbital} confirm that the system does not transit to the propeller regime at any orbital phase between type I outbursts. Such a behaviour is very similar to what was recently revealed in the another transient XRP GRO~J1008$-$57, where the luminosity stopped its fading around 10$^{35}$ erg s$^{-1}$ \citep{Tsygankov2017a}. This was interpreted as a transition of the source to the stable accretion from the `cold' low-ionization disc  \citep{Lasota2001}. 
Based on the analogy with GRO~J1008$-$57 where the magnetic field is known from the cyclotron line and using  Eq.~(6) from \citet{Tsygankov2017a} one can expect the magnetic field in \source\ to be below $\sim10^{13}$~G.

Another indirect method to evaluate the magnetic field of the NS is based on the possibility of estimating the inner radius of the accretion disc (or magnetospheric radius) from the properties of the PDS. The origin of the flux variability can be explained with the perturbation propagation model \citep{Lyubarskii1997}. In the framework of this model, stochastic viscous processes in the accretion disc perturb the mass accretion rate at a given radius on the time scales close to the local Keplerian frequency. The resulting PDS of mass accretion rate at inner disc radius can be described with the power law up to maximal frequency which can be generated in the disc \citep{Lyubarskii1997,Mushtukov2018}. In the case of XRPs, due to the strong magnetic field of the NS, the disc is truncated by the magnetosphere at the radius 
\begin{equation}
\label{eq:r_m}
R_\text{m} = 2.6\times 10^8 \xi M^{1/7} R_\text{6}^{10/7} B_{12}^{4/7} L_{37}^{-2/7}\ \text{cm},
\end{equation}
where $\xi$ is a parameter describing the accretion geometry, typically taken to be $0.5$; $M$ is the NS mass in units of solar masses; $R_\text{6}$ is the NS radius in units of $10^6$~cm; $B_{12}$ is the magnetic field strength in units of $10^{12}$~G and $L_{37}$ is the luminosity in units of $10^{37}$~erg~s$^{-1}$ \citep{Lamb1973}.

At smaller radii the noise is not generated and one can expect break, i.e. steepening of the power-law index from $\sim -1$ to $\sim -2$, in the PDS at maximal frequency $f_\text{\rm b}$ corresponding to the magnetospheric radius. 
As a first approximation, the break frequency is related to the Keplerian orbital frequency $\Omega_\text{K}$ at the innermost radius of the disc and offers a way to connect the power spectrum feature to the magnetic field \citep{Revnivtsev2009,2012MNRAS.421.2407T,2014A&A...561A..96D}: 
\begin{equation}
\label{eq:f_b}
f_\text{b} = \frac{1}{2\pi}\Omega_\text{K}(R_\text{m})/2\pi = \sqrt{\frac{G M}{ R_\text{m}^3}}  .
\end{equation}

The PDS based on the \textit{NuSTAR} data presented in Fig.~\ref{fig:pds} does not exhibit a significant break and can well be  fitted with a simple power law. Based on the broken power-law fit, we can give only a lower limit estimate to the break frequency, $f_\text{b}\gtrsim 0.6$~Hz. From Eq.\,\eqref{eq:f_b} we see that in this case the inner radius has to be smaller than $2.4\times 10^8$~cm. Assuming luminosity during our \textit{NuSTAR} observation of $5.5 \times 10^{36}$~erg~s$^{-1}$ and using Eq.\,\eqref{eq:r_m}, we can estimate an upper limit for the magnetic field $B\lesssim 2 \times 10^{12}$~G, which is within the standard range of magnetic fields for XRPs. If, however, the initial mass accretion rate fluctuations are driven by the dynamo process at frequency a few times below the local Keplerian one \citep{2004MNRAS.348..111K}, the resulting magnetic field strength might be even lower by about the same factor.


\section{Conclusions}

In this work, we presented the results of the X-ray and IR analysis of the poorly studied XRP \source\ and its optical companion during the transition from type I outburst to the quiescent state. Data obtained from the \textit{Chandra}, \textit{Swift} and \textit{NuSTAR} observatories alongside with data from UKIDSS/GPS and \textit{Spitzer}/GLIMPSE surveys were used. The aim of this study was to describe properties of the source in the hard X-ray band as well as its optical companion for the first time. To estimate the distance to the source and the nature of the companion star, we performed an analysis of IR data. As a result, it is suggested that the companion star in this system can be classified as a B0-2e star located at a distance of 7--13~kpc. We attempted to estimate the strength of the NS magnetic field using different methods. The spectral analysis revealed no cyclotron line in the  \textit{NuSTAR} broad-band spectrum, that corresponds to the magnetic field $B \lesssim 5\times10^{11}$ or  $B \gtrsim 6\times10^{12}$~G. According to the long-term light curve, the system does not switch to the propeller state, keeping a relatively high luminosity of about $10^{34}$--$10^{35}$ erg s$^{-1}$ between type I outbursts. Using similarity of this behaviour to another transient Be/XRP GRO~J1008$-$57 and based on the `cold' disc model, the magnetic field can be roughly estimated as $B \lesssim 10^{13}$~G. Another method based on the properties of the fast flux variability resulted in a lower value of an upper limit for the magnetic field strength $B \lesssim 2\times10^{12}$~G. It is important to note that such methods are applicable only in the case of accretion from the disc. More sensitive observations in the low state are required to make final conclusions.

\begin{acknowledgements}
      This study was supported by the grant TM-17-10606 of the Finnish National Agency for Education (AN), the Academy of Finland travel grants 309228, 316932 and 317552 (SST, JP) and the Russian Foundation for Basic Research projects 16-02-00294 (DIK) and 17-52-80139 BRICS-a (AAL, SST).
      This work is based in part on data of the UKIRT Infrared Deep Sky Survey. Also, part of this work is based on observations made with the Spitzer Space Telescope, which is operated by the Jet Propulsion Laboratory, California Institute of Technology under a contract with NASA. 
      We also express our thanks to the {\it NuSTAR} and {\it Swift} teams for prompt scheduling and executing our observations. 
\end{acknowledgements}

\bibliographystyle{aa} 
\bibliography{ref-final.bib} 

\end{document}